\documentclass[preprint,12pt]{aastex}
\newcommand{\bea}{\begin{eqnarray} }
\newcommand{\eea}{\end{eqnarray}}
\def\mbf#1{\mbox{\boldmath ${#1}$}}

\begin{document}

\title{The Three-Dimensional Structure of a Massive Gas Disk in 
the Galactic Central Region}

\author{Keiichi WADA}
\affil{National Astronomical Observatory of Japan, Mitaka, Tokyo
181-8588, Japan\\
E-mail: wada.keiichi@nao.ac.jp}




\begin{abstract}
Using high-resolution, three-dimensional hydrodynamical simulations, we
investigate the structure of 
the interstellar medium in the central hundred pc region in galaxies,
taking into account self-gravity of the gas,
radiative cooling from 10 K to $10^8$ K, and
energy feedback from supernovae.
Similar to the previous two-dimensional results produced 
by \citet{wad99,wad01},
we find that a gravitationally and thermally unstable ISM
evolves, in a self-stabilizing manner, into a quasi-stable thin 
disk, which is characterized by a network of cold ($T < 100$ K), 
dense clumps and filaments, and hot ($T > 10^6$ K), diffuse medium. 
Supernova explosions blow the diffuse gases from the disk, and as a result, 
a quasi-steady diffuse halo, which is not uniform but has a 
plume-like structure, is formed.
 The density probability distribution function (PDF) in a quasi-steady state
is well fitted by a Log-Normal function over about seven orders of 
magnitude.

\end{abstract}




%


\section{INTRODUCTION}

Massive gas components whose total masses are $10^7\sim 10^8 M_\odot$ 
are widely observed in the central kpc and sub-kpc
regions of various galaxies.
 Such a massive gas component plays an important role in such nuclear
activity as nuclear starbursts, galactic superwind, and AGNs.
However, the structure of the
massive gas disk, or its relation with star-formation and fueling 
processes in the AGN region, are not yet understood.
For example, the millimeter interferometers,
do not have fine enough spatial resolution 
to reveal detailed structure of the molecular gas in the central 100 pc
region of nearby galaxies.
Hopefully this situation
will drastically change in the next decade; using the next generation
millimeter and submillimeter interferometer ($ALMA$), the
spatial resolution is improved to $\sim 0.01$ arcsecond,
and we expect to be able to know pc-scale structures of the molecular gas
in the central regions of nearby galaxies.
On the other hand, our theoretical understanding of the ISM
in the central 
sub-kpc region is still insufficient. The ISM in the central region 
should not 
simply be approximated as a one-phase fluid or an ensemble of discrete clumps. 
We should use a more realistic treatment for the ISM in the central region
to reveal how the structure of the ISM relates to the nuclear activity.

Recently \citet{wad99,wad01} (hereafter WN99, WN01)
 presented high-resolution, 2D numerical
models of the ISM in a kpc-scale galactic disk, in which they numerically 
solve hydrodynamical equations and the Poisson equation
with realistic radiative cooling and heating.
Star-formation and its energy feedback on the ISM are implemented in 
their numerical code. 
They found a quasi-stable structure of the
ISM where various temperatures and density phases of the gases
coexist. Its velocity field resembles that of compressible turbulence.
However the 2D approximation used in WN01 would not
be relevant for the 3D ISM, especially for  the vertical 
structure of the hot gas component \citep{rosen95,korpi99}
in the central region of galaxies.

In this {\it Letter}, we report on a first attempt to understand
3D structure of the ISM in the galactic central region,
taking into account
self-gravity of the gas, galactic differential 
rotation, radiative cooling, and heating
due to UV background radiation and supernova (SN) explosions.

\section{NUMERICAL METHOD AND MODELS}

The numerical methods are basically the same as those described in WN01.
Here we briefly summarize them.
\setcounter{footnote}{0} We solve the following equations numerically
in three dimensions to simulate the evolution of a rotating ISM in 
a fixed gravitational potential.
  \begin{eqnarray}
\frac{\partial \rho}{\partial t} + \nabla \cdot (\rho \mbf{v}) &=& 0,
\label{eqn: rho} \\ \frac{\partial \mbf{v}}{\partial t} + (\mbf{v}
\cdot \nabla)\mbf{v} +\frac{\nabla p}{\rho} + \nabla \Phi_{\rm ext} +
\nabla \Phi_{\rm sg} &=& 0, \label{eqn: rhov}\\
 \frac{\partial E}{\partial t} + \frac{1}{\rho} \nabla \cdot 
[(\rho E+p)\mbf{v}] &=& 
\Gamma_{\rm UV} + \Gamma_\star- \rho \Lambda(T_g), \label{eqn: en}\\ \nabla^2
\Phi_{\rm sg} &=& 4 \pi G \rho, \label{eqn: poi} 
\end{eqnarray}

 where, $\rho,p,\mbf{v}$ are the density, pressure, and velocity of
the gas, and the specific total energy $E \equiv |\mbf{v}|^2/2+
p/(\gamma -1)\rho$, with $\gamma= 5/3$.  We assume a time-independent
external potential $\Phi_{\rm ext} \equiv  -(27/4)^{1/2}v_c^2/(r^2+
a^2)^{1/2}$, where $a = 10$ pc is the core radius of the potential and $v_c = 100$ km s$^{-1}$ is the maximum rotational velocity.  We also assume a cooling function
$\Lambda(T_g) $ $(10 < T_g < 10^8 {\rm K})$ \citep{SN} with Solar metallicity
and a heating due to photoelectric heating, $\Gamma_{\rm UV}$ 
and due to energy feedback from SNe, $\Gamma_\star$.
We assume a uniform UV radiation field,
which is ten times larger than the local UV field \citep{ger97}:
$ \Gamma_{\rm UV} = 1.0 \times 10^{-23} \varepsilon   
G_0 \, {\rm ergs \: s}^{-1}, $
where the heating efficiency $\varepsilon$ is assumed to be 0.05 and
$G_0$ is the incident FUV field normalized to the local interstellar
value.

The hydrodynamic part of the basic equations is solved by AUSM
 (Advection Upstream Splitting Method), \citep{LS}. 
We use $512^2 \times 32$ Cartesian grid points covering a $256^2 \times
16$ pc$^3$ region around the galactic center.
Therefore, the spatial resolution is 0.5 pc. 
The Poisson equation is solved to calculate self-gravity of the gas
using the FFT and the convolution method. 
The second-order leap-frog method is used for the time integration.  We
adopt implicit time integration for the cooling term.
The initial condition is an axisymmetric and rotationally supported
thin disk of a uniform density profile (thickness is 2.5 pc) with a total gas mass of  $M_g = 5\times 10^7 M_\odot$
(initial average density is about $4\times 10^4$ cm$^{-3}$).
Random density and temperature fluctuations are added to the
initial disk. These fluctuations are less than 1 \% of the unperturbed
values and have an approximately white noise distribution. The initial
temperature is set to $10^4$ K over the whole region. 

SN explosions are assumed to occur at random positions
 in the region of 
$|x|,|y| < 115$ pc and $|z| < 4$ pc. The average SN
rate is 4.8 yr$^{-1}$ kpc$^{-2}$,
which is as high as that in nuclear starbursts \citep{ken98}.
The energy of $10^{51}$ ergs is instantaneously
injected into a single cell as thermal energy. 
During the calculation ($\sim 5$ Myr), about $10^5$ SNe explodes.
We do not assume simple evolutionary models
for each SNR or for the heating efficiency of the ISM due to SNe.
The 3D evolution of blast waves caused by SNe in an inhomogeneous and non-stationary medium with global
rotation is followed explicitly, taking into account the radiative cooling.
Therefore the evolution of the SNRs, e.g. the duration and structure 
of the SNe, depends on the gas density distribution around the SN.

Computational time for one typical run is about 30 hours using 
a vector parallel supercomputer (9.6 Gflops/processor $\times$ 32 processors).
%
\section{RESULTS}
%
Figure 1 (a)  shows 
the 3D density distribution of the disk in a
quasi-stable state ($t=4.3$ Myr).
Dense filaments/clumps form a ``tangled network''. 
High density clumps colored yellow are embedded in 
less dense filaments colored blue. Low density regions occupy 
a larger volume than the dense components. 
The morphology does not differ much from that found in 2D
global models (WN01). The formation process of the complicated structure is 
the same as in the 2D case (see Fig. 6 in WN01). 
A number of processes are involved in the
formation of the filamentary structure: 1) tidal and collisional 
interactions between dense regions formed due to
thermal and gravitational instabilities, 2) differential rotation, 3) shear motion due to local turbulent 
motion, and 4) SN explosions. 
The last mechanism is not a dominant process in the disk plane, where
the radiative cooling is very effective due to the high gas density,
and most of the SN energy is used to form vertical structure (see below).
Radial surface density distribution in the quasi-steady state is
fitted on average by an exponential profile: 
$\Sigma_g (R) \sim 200 e^{-R/R_0} M_\odot$ pc$^{-2}$, 
with $R_0 \sim 50$ pc. This is a consequence of the angular momentum transport
due to the turbulent motion in the disk. Velocity dispersion in the plane is 
$\sim 35\pm 5$ km s$^{-1}$ over the whole disk.
Figure 1 (b) is an edge-on view of 
the density distribution.
Vertical distributions of average density and temperature are also
shown. The dense, cold gases form a geometrically 
thin disk, where the average temperature is about $5000$ K.
Note that most of the volume is occupied by cold ($T_g < 100 $ K) gas
near the disk plane  (see Fig. 2(a)),
but warm and hot regions ($T_g > 10^5$) increase the average
temperature. 
The dense disk has a diffuse envelope, where 
the volume average temperature is about $3-4 \times 10^4$ K.
The diffuse gases form plume-like structures outside the dense thin disk,
 which are direct results of SN explosions occurring near 
the disk plane. The vertical structure is quasi-steady in a global sense.
That is, the kinetic and thermal heating due to SNe
balances with the radiative cooling followed by infall of the cooled gas
towards the disk.
It looks like a miniature of the ``galactic fountain'' model \citep{SF76, IH80}.

It would be useful to estimate a typical radius of SNRs for understanding 
what causes the density structures.
A radiative SNR becomes in a pressure-driven snowplow (PDS) phase when the
radius of the SNR is 
$R_s \sim 0.6 (\rho/100 M_\odot$ pc$^{-3})^{-1/2}$ pc \citep{cio91}.
 The maximum radius of the shell at the end of the snowplow phase
 is $R_{s,{\rm max}} \sim 80 \rho^{-1/5} P_4^{-1/5}$ pc, where the ambient pressure $P_4 \equiv P/(k_B \times 10^4$ cm$^{-3}$ K). 
The velocity field of the medium is turbulent-like due to 
the random energy input from SN explosions, and the vertical rms velocity
is $ v_{\rm rms} \sim 10$ km s$^{-1}$, which is comparable to that estimated from 
the evolution of shock velocity of a SNR ($\propto t^{-3/5}$) after
1 Myr evolution.
If we use the turbulent pressure $\rho v_{\rm rms}^2$ as $P_4$,
$R_{s,{\rm max}} \sim 3$ pc. Since $R_s \propto t^{2/7}$ at the PDS phase, the remnant reaches its maximum size at $t\sim 1$ Myr.
Above estimate implies that the SNRs can not 
be larger than 1 pc near the disk plane where the average
density is much greater than 100 $M_\odot$ pc$^{-3}$.
SNe that explodes at several pc above the disk
plane may cause 10 pc-scale {\it vertical} structures as seen in Fig. 1(b). 
Note that the present spatial resolution (0.5 pc) might not be fine enough to
resolve the early evolution of SNRs in very dense regions. 

Figure 2 is the temperature distribution (a) at the
disk plane ($z = 0.$), and (b) at $z = 4$ pc  at $t = 5.2$ Myr. 
In the disk plane, most gases are in a cold ($T_g < 100$ K),
while on the other hand, 
warm ($ 10^5 > T_g > 10^2$ K) and hot phases ($T_g > 10^5$ K) 
create a patchy morphology above the disk plane. 
The hot gas originates in SN explosions.
The UV radiation contributes to form the warm, diffuse gas, but it does not
significantly affect the gas dynamics.
Since the radiative cooling in the
dense regions near $z = 0$ is effective, the SNRs are less prominent
near the disk plane 
than in the halo region.

Although the spatial structure of the density is 
quite complicated as seen in Fig. 1, statistically the system is rather simple.
In Fig. 3, we plot the probability distribution function of the density,
i.e. a histogram of volume as a function of the gas density.
The PDF is well fitted by a single Log-Normal function 
over about seven orders of magnitude between
$\rho = 10^{-2} \sim 10^6 M_\odot$ pc$^{-3}$,
which suggests strong non-linearity in the formation process.

Figure 4 is the volume-weighted temperature PDF.
There are two dominant phases  at $T_g \sim 50$ K and $10^4$ K, which 
are thermally stable phases determined by the cooling functions 
(see Fig. 1 in WN01). The cold gas around $T_g \sim 50$ K corresponds
to the point $d \log \Lambda /d \log (T_g) \sim  -2 $, above which
the system is expected to be thermally unstable \citep{SM72}. 
There are also small peaks at $T_g \sim 7\times 10^4$ and $\sim 10^6$ K. 
The latter is a direct result of SN explosions.
The system roughly shows three phases in temperature.
It should be noted, however, that a considerable amount  
gas exists between the cold and warm phases which are
thermally unstable regimes. This cannot be understood by the behavior of 
multi-phase gas in pressure equilibrium,
 but as a consequence of turbulent motions in the medium.
As in the 2D case (see Fig. 13 in WN01), the thermal
pressure is dominated by turbulent pressure in this system.
\citet{VZ00} concluded from their 2D simulations that
the turbulence smears out the thermal phases, creating a continuous
distribution of the physical properties. This is consistent with 
our 3D results.

%
\section{DISCUSSION}
%
We have examined the 3D density and temperature 
structure of a massive gas component in the central 100 pc region of a galaxy.
We found that a globally stable, multi-phase disk is formed as a
natural consequence of non-linear evolution.
The structure is similar to what has previously been found in
2D simulations for a kpc-scale disk (WN99, WN01).
The SN explosions are assumed 
to be as high as those in nuclear starbursts, 
but they do not drastically change the intrinsically inhomogeneous
``tangled'' network density structure near the disk plane ($z = 0$).
However, the SNe are important in producing 
 the vertical structure, i.e.
the diffuse filamentary halo ($\rho \sim 10-10^{-2} M_\odot$ pc$^{-3}$), and
the hot component ($T_g > 10^6$ K).
We also found that
the density probability distribution function (PDF) is well fitted by
a single Log-Normal function over about seven orders of magnitude.
This suggests that the density structure
is dominated by a highly non-linear process from low to high density regimes
\citep{VZ94}.

Probability density functions (PDFs) have been discussed in relation
to the structure of the turbulent interstellar medium (ISM). 
\citet{NP} indicated a formal proof for the Log-Normal PDF in
isothermal supersonic turbulence. 
They also showed numerically that the
Log-Normal PDF appears in 3D isothermal, non-self-gravitating
 supersonic turbulence over about three orders of 
magnitude.
In our system, on the other hand, thermal and gravitational instabilities
play an important role to produce the very dense regions.
The PDF can be fitted by a Log-Normal function over seven orders of 
magnitude. Therefore, we believe that the isothermality is not a necessary
condition for the Log-Normal PDF, 
but it is a natural feature of 
the highly non-linear and random processes, which are achieved 
in the system that we are investigating, 
i.e., the self-gravitating, supersonic turbulence with radiative cooling
and stellar energy feedback. However, an important open question remains:
What determines the ``width'' of the Log-Normal PDF? \citet{NP} suggested that
the standard deviation of the Log-Normal PDF measures the mean rate of change of the density or r.m.s. Mach number of the turbulence.
Since the density range in which the Log-Normal PDF is achieved in our system
is quite large, and self-gravity of the gas is crucial for yielding the
very high density regions, it is not likely that the r.m.s. Mach number 
is the only parameter needed to determine the width of the Log-Normal PDF.
WN01 reported that the density PDF in 2D
models can be fitted with a Log-Normal function for a high density region 
over about four
 orders of magnitude, but that the low density part 
is fitted by a Normal function.
In the present 3-D model, 
 we do not find that the PDF for a low density region obeys the Normal distribution.
The difference between the 2D and 3D results
would suggest that the third dimension is more important for less dense
gases.

Finally, the density structure of the massive disk with SNe 
is interesting in terms of 
the starburst-AGN connection. If the gas system around the AGN 
is inhomogeneous as seen in the present model,
then the nucleus could be obscured by the dense clumps and filaments
from a certain line-of-sight. The turbulent motion in the disk causes 
mass inflow towards the nucleus, and the accretion rate would be
time-dependent, and would also depend on the star formation rate in the disk
(Wada \& Norman, in preparation).

\acknowledgments I am grateful to Colin Norman for many stimulating 
discussions.
Numerical computations were carried out on Fujitsu VPP5000 at NAOJ.  
This work is supported in part by Grant-in-Aids for Scientific Research 
(no. 12740128) of JSPS.
\clearpage

\newpage
\figcaption{(a) 3D density distribution in a quasi-stable 
state visualized by 
the volume-rendering technique. Dense clumps which are represented 
by yellow are embedded in less dense filaments colored 
green or blue. (b) Same as Fig. 1(a), but for the edge-on 
view. Diffuse plumes are formed outside the dense, geometrically thin disk. 
Average density and temperature distributions for the $z$-direction are
plotted with filled and open circles, respectively. The units are
$M_\odot$ pc$^{-3}$ for density and $K$ for temperature.}
\figcaption{(a) Temperature distribution at the mid-plane ($z=0$) at $t=5.2$ Myr. The color bar shows the temperature range 
between 10 to $10^6$ K in a logarithmic scale.
The red regions are young SN remnants. 
(b) Same as (a), but for a slice at $z=4$ pc. Hot ($T_g > 10^6$ K) and warm 
($T_g \sim 10^2-10^5$ K) gas are more dominant than at the disk plane.}
\figcaption{Probability density function (PDF) at $t=5.18$ Myr. 
The PDF is fitted by a single Log-Normal function; its average density
is $10^{-0.5} M_\odot$ pc$^{-3}$ and the dispersion is $10^{2.1}  M_\odot$ pc$^{-3}$. }
\figcaption{Volume-weighted
 temperature distribution at $t=5.18$ Myr. $N_0$ means
the total number of zones.}

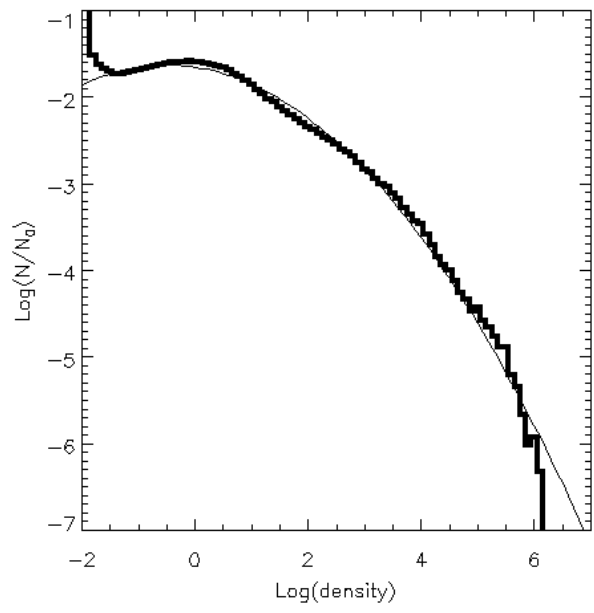
\end{document}